\documentclass[12pt]{iopart}
\usepackage{iopams}
\usepackage{amsthm}
\usepackage{graphicx}
\newtheorem{Thm}{Theorem}

\newtheorem{Lem}{Lemma}

\newcommand{\bra}[1]{{\left\langle #1 \right|}}
\newcommand{\ket}[1]{{\left| #1 \right\rangle}}

\newcommand{\C}{\mbox{$\mathbb C$}}

\newcommand{\T}{\mbox{$\mathrm{tr}$}}

\begin{document}
\title{Generalized W-Class State and its Monogamy Relation}

\author{Jeong San Kim and Barry C. Sanders }

\address{
 Institute for Quantum Information Science,
 University of Calgary, Alberta T2N 1N4, Canada
}
\eads
{\mailto{jkim@qis.ucalgary.ca},
\mailto{bsanders@qis.ucalgary.ca}}

\date{\today}

\begin{abstract}
We generalize the W class of states from $n$ qubits to $n$ qudits and prove that their entanglement
is fully characterized by their partial entanglements even for the case of the mixture that consists of
a W-class state and a product state $\ket{0}^{\otimes n}$.
\end{abstract}

\pacs{
03.67.-a, 
03.65.Ud, 
}
\maketitle

\section{Introduction}

Quantum entanglement is one of the most non-classical features in quantum mechanics and provides us
a lot of applications. Due to its variety of usages, much attention has been shown, so far, for
the quantification of entanglement and, thus, the concept of entanglement measure has been naturally arisen.
{\em Concurrence}~\cite{Wootters} is one of the most well-known bipartite entanglement measures with
an explicit formula for $2$-qubit system while there does not exist any analytic way of
evaluation yet for the general case of higher-dimensional mixed states.
Another entanglement measure that can be considered as a dual to concurrence is the
{\em concurrence of assistance} (CoA)~\cite{LVE}, and this can be interpreted as the
maximal average concurrence that two parties in the bipartite system can locally prepare with the help of the third party who has
the purification of the bipartite system.

In multipartite quantum system, there can be several inequivalent types of entanglement among the subsystems and
the amount of entanglement with different types might not be directly comparable to each other.
 For $3$-qubit pure states, it is known that there are
two inequivalent classes of genuine tripartite entangled states~\cite{DVC}; one is
the Greenberger-Horne-Zeilinger (GHZ) class~\cite{GHZ},
and the other one is the W-class~\cite{DVC}.
This can be characterized by means of stochastic local operations and classical communication (SLOCC),
that is, the conversion of the states in a same class can be achieved through local operation and classical communication
with non-zero probability.

Another way to characterize the different types of entanglement distribution is by using {\em monogamy relation} of entanglement.
Unlike classical correlations, the amount of entanglement that can be shared between any of two parties and the others
is strongly constrained by the entanglement between two parties.

In $3$-qubit systems, Coffman, Kundu and Wootters (CKW)~\cite{CKW} first introduced a monogamy inequality in terms of a bipartite
entanglement measure, concurrence, as
\begin{equation}
\mathcal{C}_{A(BC)}^2\ge\mathcal{C}_{AB}^2+\mathcal{C}_{AC}^2,
\label{CKW}
\end{equation}
where $\mathcal{C}_{A(BC)}=\mathcal{C}(\ket{\psi}_{A(BC)})$ is the concurrence of a 3-qubit state $\ket{\psi}_{A(BC)}$ for a bipartite cut of subsystems between
 $A$ and $BC$ and $\mathcal{C}_{AB}=\mathcal{C}(\rho_{AB})$. Its generalization into
  $n$-qubit case was also proved ~\cite{OV} and, symmetrically,
 its dual inequality in terms of the CoA for 3-qubit states,
\begin{equation}
\mathcal{C}_{A(BC)}^2\le(\mathcal{C}_{AB}^a)^2+(\mathcal{C}_{AC}^a)^2,
\label{dual}
\end{equation}
and its generalization into $n$-qubit cases have been shown in~\cite{GMS,GBS}.

In $3$-qubit system, two inequivalent classes of genuine tripartite entangled
states, GHZ and W classes, show extreme difference in terms of CKW inequality and its dual one.
In other words, CKW inequality is saturated by W-class states,
while it becomes the most strict inequality with the states in GHZ class.
Here, W-class states are of our special interest, since the saturation of the inequality implies that
a genuine tripartite entanglement can have a complete characterization by means of the bipartite ones inside it.
In other words, the entanglement $A$-$BC$, measured by concurrence, is completely determined by its partial entanglements, $A$-$B$ and $A$-$C$.
For the case of $n$-qubit W-class states, generalized CKW and its dual inequalities are also saturated, and
thus the same interpretation can be applied.

In this paper, we generalized the concept of W-class states from $n$ qubits to $n$ qudits and show that their entanglement
is fully characterized by their partial entanglements. We also prove that the complete characterization of the global
entanglement in terms of its partial entanglement is possible even for the case of the mixture consisting of
a W-class state and a product state $\ket{0}^{\otimes n}$.

This paper is organized as follows. In Section~\ref{Sec:nW}
we recall the the monogamy relation of $n$-qubit W-class states in terms of CKW and its dual inequalities.
In Section~\ref{Structure}, we provide more general monogamy relations of $n$-qubit W-class states with respect to
arbitrary partitions by investigating the structure of $n$-qubit W-class states.
In Section ~\ref{Subsec:ndW} we generalize the concept of W-class states to arbitrary $n$-qudit system as well as
its monogamy relations with respect to arbitrary partitions.
In Section~\ref{Sec:mixed}, we consider the class of multipartite mixed states that is a mixture of a W-class state and a product state.
We also provide its monogamy relations in terms of its partial entanglement by studying its structural properties. In Section~\ref{Sec:conclusion}
we summarize our results.
\section{Monogamy Relation of $n$-qubit W-class States}\label{Sec:nW}
For any bipartite pure state $\ket{\phi}_{AB}~\in \mathbb{C}^d \otimes \mathbb{C}^{d'}$,
concurrence of $\ket{\phi}_{AB}$ is defined as
\begin{equation}
\mathcal{C}(\ket{\phi}_{AB})=\sqrt{2(1-\T\rho_A^2)},
\label{c_pure}
\end{equation}
where $\rho_A=\T_B\ket{\phi}_{AB}\bra{\phi}$.
For any mixed state $\rho_{AB}$,
it is defined as
\begin{equation}
\mathcal{C}(\rho_{AB})=\min \sum_{k} p_k \mathcal{C}(\ket{\phi_k}_{AB}),
\label{eq:concurrence}
\end{equation}
where the minimum is taken over its all possible pure state decompositions,
$\rho_{AB}=\sum_k p_k \ket{\phi_k}_{AB}\bra{\phi_k}$.

Another entanglement measure that can be considered as a dual to concurrence is CoA~\cite{LVE},
which is defined as
\begin{equation}
\mathcal{C}^a(\rho_{AB})=\max \sum_{k} p_k \mathcal{C}(\ket{\phi_k}_{AB}),
\label{eq:CoA}
\end{equation}
where the maximum is taken over all possible decompositions of $\rho_{AB}$.

For $3$-qubit W-class states
\begin{equation}
\ket{W}_{ABC} = a\ket{100}_{ABC}+b\ket{010}_{ABC}+c\ket{001}_{ABC},
\label{3W}
\end{equation}
with $|a|^2 +|b|^2 +|c|^2 =1$, CKW and its dual inequalities (\ref{CKW}), (\ref{dual}) are saturated, that is,
\begin{eqnarray}
\mathcal{C}_{A(BC)}^2 &= \mathcal{C}_{AB}^2 + \mathcal{C}_{AC}^2,\nonumber\\
\mathcal{C}_{AB} &= \mathcal{C}_{AB}^a,~\mathcal{C}_{AC}= \mathcal{C}_{AC}^a .
\label{3satu}
\end{eqnarray}

In other words, the entanglement of W-class states between one party and the rest can have a complete characterization
 in terms of the $\emph{partial entanglement}$ that is the bipartite entanglement between one party and each of the rest parties.

For $n$-qubit systems $A_1 \otimes \cdots \otimes A_n$ where $A_i \cong \mathbb{C}^2$ for $i=1,\ldots,n$,
 CKW and its dual inequalities can be generalized as~\cite{OV,GBS}
\begin{eqnarray}
\mathcal{C}_{A_1 (A_2 \cdots A_n)}^2 & \geq & \mathcal{C}_{A_1 A_2}^2 +\cdots+\mathcal{C}_{A_1 A_n}^2 ,
\nonumber\\
\mathcal{C}_{A_1 (A_2 \cdots A_n)}^2 & \leq & (\mathcal{C}^a_{A_1 A_2})^2 +\cdots+(\mathcal{C}^a_{A_1 A_n})^2 .
\label{nCKW}
\end{eqnarray}
For $n$-qubit W-class states,
\begin{eqnarray}
\ket{W}_{A_1\cdots A_n} =a_1 \ket{1\cdots0}_{A_1\cdots A_n}+\cdots+a_{n} \ket{0\cdots1}_{A_1\cdots A_n},\nonumber\\
\sum_{i=1}^{n}|a_i|^2 =1,
\label{nWclass}
\end{eqnarray}
the inequalities (\ref{nCKW}) are saturated; that is,
\begin{eqnarray}
\mathcal{C}_{A_1 (A_2 \cdots A_n)}^2 = \mathcal{C}_{A_1 A_2}^2+\cdots+\mathcal{C}_{A_1 A_n}^2 ,
\nonumber\\
\mathcal{C}_{A_1 A_i} = \mathcal{C}_{A_1 A_i}^a,~~i=2,\ldots,n.
\label{nsatu}
\end{eqnarray}

In fact, there can be several ways to show Equation~(\ref{nsatu}). Since any two-qubit reduced density matrix $\rho_{A_1 A_i}$
of $\ket{W}_{A_1 \cdots A_n}$ can have analytic formulas for concurrence and concurrence of assistance~\cite{Wootters,LVE},
one of the ways to check Equalities (\ref{nsatu}) is using the formulas in~\cite{Wootters, LVE}.
However, there is no formula for the general case of a bipartite quantum state with arbitrary dimension,
so here we use the method of considering all possible decompositions of the bipartite mixed state $\rho_{A_1 A_i}$ so that
the optimization process in (\ref{eq:concurrence},~\ref{eq:CoA}) can also be used later in
this paper for higher-dimensional quantum systems.

Let us first consider $\mathcal{C}^2_{A_1 (A_2 \cdots
A_n)}=2(1-\T\rho_{A_1}^2)$ where $\rho_{A_1}=\T_{A_2\cdots
A_n}({\ket{W}_{A_1\cdots A_n}\bra{W}})$.
Since $\rho_{A_1}=|a_1|^2\ket{1}_{A_1}\bra{1}+(\sum_{i=2}^{n}|a_i|^2)\ket{0}_{A_1}\bra{0}$, we can easily see
\begin{equation}
\mathcal{C}^2_{A_1 (A_2 \cdots A_n)}= 4|a_1|^2(\sum_{i=2}^{n}|a_i|^2).
\label{C_1(2_n)}
\end{equation}
For $\mathcal{C}^2_{A_1 A_i}$ with $i \in \{2,\ldots,n\}$, let us consider
\begin{eqnarray}
\rho_{A_1 A_i} &=&\T_{A_2\cdots\widehat{A}_i\cdots A_n}({\ket{W}_{A_1\cdots A_n}\bra{W}})\nonumber\\
               &=&(a_1\ket{10}+a_i\ket{01})_{A_1 A_i}(a_1^*\bra{10}+a_i^*\bra{01})\nonumber\\
               &&+(|a_2|^2 +\cdots+|\widehat{a}_i|^2+\cdots+|a_n|^2)\ket{00}_{A_1 A_i}\bra{00},
\label{rhoA1Ai}
\end{eqnarray}
where $A_2\cdots \widehat{A}_i\cdots A_n = A_2\cdots A_{i-1}A_{i+1}\cdots A_n$ and
 $(|a_2|^2 +\cdots+|\widehat{a}_i|^2+\cdots +|a_n|^2)=(|a_2|^2 +\cdots+|a_{i-1}|^2+|a_{i+1}|^2+\cdots+|a_n|^2)$.
Now, let $\ket{\tilde{x}}_{A_1 A_i}=a_1\ket{10}_{A_1 A_i}+a_i\ket{01}_{A_1 A_i}$ and $\ket{\tilde{y}}_{A_1 A_i}=\sqrt{|a_2|^2 +\cdots+|\widehat{a}_i|^2+\cdots|a_n|^2}\ket{00}_{A_1 A_i}$,
where $\ket{\tilde{x}}_{A_1 A_i}$ and $\ket{\tilde{y}}_{A_1 A_i}$ are unnormalized states of the subsystems $A_1 A_i$.
Then, by the Hughston-Jozsa-Wootters (HJW) theorem~\cite{HJW}, any other decomposition of
$\rho_{A_1 A_i}=\sum_{h=1}^{r}|\tilde{\phi}_h\rangle_{A_1 A_i} \langle\tilde{\phi}_h|$ with size $r \geq 2$
can be obtained by an $r\times r$ unitary matrix $(u_{hl})$ where
\begin{equation}
|\tilde{\phi}_h\rangle_{A_1 A_i}=u_{h1}\ket{\tilde{x}}_{A_1 A_i}+u_{h2}\ket{\tilde{y}}_{A_1 A_i}.
\label{decomp}
\end{equation}
Let $\rho_{A_1 A_i}=\sum_{h}{ p_h \ket{\phi_h}_{A_1 A_i}\bra{\phi_h}}$
where $\sqrt{p_h}\ket{\phi_h}_{A_1 A_i}=|\tilde{\phi}_h\rangle_{A_1 A_i}$ and $ p_h =|\langle\tilde{\phi}_h|\tilde{\phi}_h\rangle|$;
then, by straightforward calculation, we can easily see that the average concurrence of the pure state
decomposition $\rho_{A_1 A_i}=\sum_{h}{ p_h \ket{\phi_h}_{A_1 A_i}\bra{\phi_h}}$ is,
\begin{eqnarray}
\sum_{h=1}^{r} p_h \mathcal{C}(\ket{\phi_h}_{A_1 A_i}) &=& \sum_{h=1}^{r}p_h \sqrt{2(1-\T{(\rho^h_{A_1})}^2)}\nonumber\\
 &=& 2|a_1||a_i|,
\label{C1i}
\end{eqnarray}
where $\rho^h_{A_1} = \T_{A_i}(\ket{\phi_h}_{A_1 A_i}\bra{\phi_h})$.
In other words, the average concurrence remains the same for any pure state decomposition of $\rho_{A_1 A_i}$, and
thus, we can have,
\begin{eqnarray}
\min \sum_{h} p_h \mathcal{C}(\ket{\phi_h}_{A_1 A_i}) = \max \sum_{h} p_h \mathcal{C}(\ket{\phi_h}_{A_1 A_i})
\label{avea1aisame}
\end{eqnarray}
where the maximum and minimum are taken over all possible decomposition of $\rho_{A_1 A_2}$.
This implies
\begin{equation}
\mathcal{C}_{A_1 A_i} = \mathcal{C}_{A_1 A_i}^a = 2|a_1||a_i|,
\label{Ca1aisame}
\end{equation}
which leads to (\ref{nsatu}).

\section{General Monogamy Relation of Multipartite W-Class States}\label{Sec:Gmonogamy}
In Section~\ref{Sec:nW}, we have seen the monogamy relations of $n$-qubit W-class states between subsystems
in terms of CKW and its dual inequalities.
Here, we investigate the structure of W-class states in $n$-qubit system
by considering arbitrary partitions of subsystems and derive more general concept of monogamy relations
between the parties.
Furthermore, we generalize the concept of W-class states to arbitrary $n$-qudit system and
also consider the monogamy relations in terms of arbitrary partitions.

\subsection{Structure of n-qubit W-class States}\label{Structure}

For $n$-qubit W-class states,
\begin{eqnarray}
\ket{W}_{A_1\cdots A_n} =a_0 \ket{0\cdots1}_{A_1\cdots A_n}+\cdots+a_{n-1} \ket{1\cdots0}_{A_1\cdots A_n},\nonumber\\
\sum_{i=0}^{n-1}|a_i|^2 =1,
\label{nWclass2}
\end{eqnarray}
let us consider a partition $P=\{P_1,\ldots,P_m \}$, $m\leq n$ for the set of subsystems $S=\{A_1,\ldots,A_n \}$  where
each of $P_s$ contains several qubits, that is,
\begin{eqnarray}
P_s=\{ A_{i_j} \},j=1,\ldots,n_s,~~ \sum_s n_s = n, \nonumber\\
P_s \cap P_t = \emptyset~\mathrm{for}~s\neq t,
~~ \bigcup_s P_s = S.
\label{partition}
\end{eqnarray}

\begin{figure}
  \hspace{1.5cm}
  \includegraphics[width=12cm]{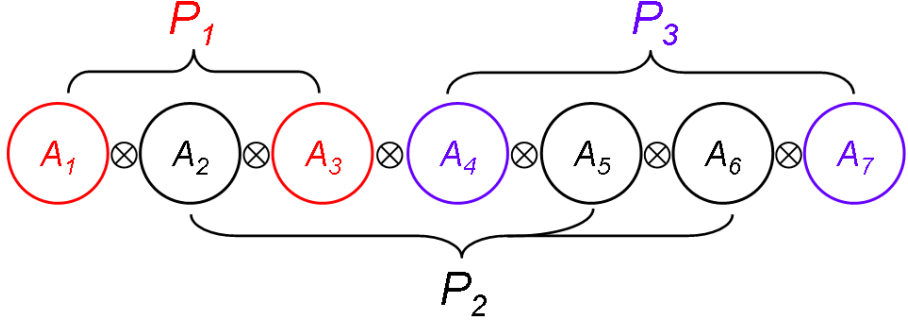}\\
  \hspace{-5cm} \caption{A partition for the set of subsystem $S=\{A_1,\ldots A_7 \}$ where $P_1 = \{A_1 , A_3\}$, $P_2 =\{A_2 , A_5 , A_6\}$
   and $P_3=\{A_4, A_7\}$.}\label{partition}
\end{figure}

For simplicity, let us first consider the case when $P=\{P_1, P_2, P_3 \}$ with $|P_s| = n_s$
 where $s \in \{1, 2, 3 \}$.
 Figure 1 shows an example of partition for a set of subsystems $S=\{A_1,\ldots A_7 \}$ where $P_1 = \{A_1 , A_3\}$, $P_2 =\{A_2 , A_5 , A_6\}$
   and $P_3=\{A_4, A_7\}$.

Without loss of generality,
we may assume $P_1 =\{A_1,\ldots,A_{n_1}\}$, $P_2 = \{A_{n_1 +1 },\ldots,A_{n_1 +n_2} \}$ and $P_2 = \{A_{n_1 + n_2 +1 },\ldots,A_{n} \}$;
 otherwise we can have some proper reordering of the subsystems.

Here, we use the representation, that is,
\begin{equation}
\ket{0\cdots1\cdots0}_{P_s}= |2 ^{i} \rangle _{P_s}
\label{decimal}
\end{equation}
where $\ket{0\cdots1\cdots0}_{P_s}$ is an $n_s$-qubit product state of the party $P_s$ whose $i$th subsystem from the right is $1$ and $0$ elsewhere.
Then (\ref{nWclass}) can be rewritten as,
\begin{eqnarray}
\ket{W}_{P_1 P_2 P_3} &=& \ket{\tilde{x}}_{P_1}|\vec{0}\rangle_{P_2}|\vec{0}\rangle_{P_3}+ |\vec{0}\rangle_{P_1}\ket{\tilde{y}}_{P_2}|\vec{0}\rangle_{P_3}\nonumber\\
                         &&+|\vec{0}\rangle_{P_1}|\vec{0}\rangle_{P_2} \ket{\tilde{z}}_{P_3},
\label{3partW}
\end{eqnarray}
where $|\vec{0}\rangle_{P_s} = \ket{0\cdots0}_{P_s}$ and $\ket{\tilde{x}}_{P_1}$, $\ket{\tilde{y}}_{P_2}$ and $\ket{\tilde{z}}_{P_3}$
 are unnormalized states in $P_1$, $P_2$ and $P_3$ respectively such that
$\ket{\tilde{x}}_{P_1}=\sum_{j=0}^{n_1 -1} a_{n_3 + n_2 + j} | 2^j \rangle_{P_1}$, $\ket{\tilde{y}}_{P_2}=\sum_{k=0}^{n_2 -1} a_{n_3 + k} |2^k \rangle_{P_2}$ and $\ket{\tilde{z}}_{P_3}=\sum_{l=0}^{n_3 -1} a_l | 2^l \rangle_{P_3}$.

Here, we note that $\ket{\tilde{x}}_{P_1}$, $\ket{\tilde{y}}_{P_2}$ and $\ket{\tilde{z}}_{P_3}$ are unnormalized W-class states of the parties
$P_1 , P_2$ and $P_3$ respectively.
Thus any $n$-qubit W-class  state can have this type of representation, that is,
\begin{eqnarray}
\ket{W}_{P_1 P_2 P_3}&=&\sqrt{q_1}\ket{W}_{P_1}|\vec{0}\rangle_{P_2}|\vec{0}\rangle_{P_3} + \sqrt{q_2}|\vec{0}\rangle_{P_1}\ket{W}_{P_2}|\vec{0}\rangle_{P_3}\nonumber\\
                         &&+\sqrt{q_3}|\vec{0}\rangle_{P_1}|\vec{0}\rangle_{P_2} \ket{W}_{P_3},
\label{3partW2}
\end{eqnarray}
where $q_1 =\sum_{j=0}^{n_1 -1}|a_{n_3 + n_2 + j}|^2$, $q_2 =\sum_{k=0}^{n_2 -1}|a_{n_3 + k}|^2$ and $q_3 =\sum_{l=0}^{n_3 -1}|a_l|^2$
 with the nomalization condition, $q_1 + q_2 + q_3 =1$.

If we just rename $\ket{W}_{P_s}=\ket{1}_{P_s}$ and $|\vec{0}\rangle_{P_s} = \ket{0}_{P_s}$, then  $\ket{1}_{P_s}$ and $\ket{0}_{P_s}$
are orthogonal to each other, and (\ref{3partW2}) can be rewritten as,
\begin{eqnarray}
\ket{W}_{P_1 P_2 P_3} &=& \sqrt{q_1}\ket{1}_{P_1} \ket{0}_{P_2} \ket{0}_{P_3} + \sqrt{q_2}\ket{0}_{P_1}\ket{1}_{P_2}\ket{0}_{P_3}\nonumber\\
                      &&+ \sqrt{q_3}\ket{0}_{P_1} \ket{0}_{P_2} \ket{1}_{P_3},
\label{3partW3}
\end{eqnarray}
which is a tripartite W-class state in ${(\mathbb{C}^2)}^{n_1}\otimes{(\mathbb{C}^2)}^{n_2}\otimes{(\mathbb{C}^2)}^{n_3}$ quantum systems.

Similarly, for an arbitrary partition $P=\{P_1,\ldots,P_m \}$ of size $m$,
we can have
\begin{eqnarray}
\ket{W}_{P_1 \cdots P_m}
 &=& \sqrt{q_1}\ket{W}_{P_1} \ket{0}_{P_2}\cdots \ket{0}_{P_m} + \sqrt{q_2}\ket{0}_{P_1}\ket{W}_{P_2}\cdots\ket{0}_{P_3}\nonumber\\
                         &&+ \cdots+ \sqrt{q_m}\ket{0}_{P_1} \ket{0}_{P_2}\cdots \ket{W}_{P_m}\nonumber\\
&=& \sqrt{q_1}\ket{1}_{P_1} \ket{0}_{P_2}\cdots \ket{0}_{P_m} + \sqrt{q_2}\ket{0}_{P_1}\ket{1}_{P_2}\cdots\ket{0}_{P_3}\nonumber\\
                         &&+ \cdots+ \sqrt{q_m}\ket{0}_{P_1} \ket{0}_{P_2}\cdots \ket{1}_{P_m}\label{2WinW}.
\label{mpartW}
\end{eqnarray}
For any partition $P=\{P_1,\ldots,P_m \}$ of the set of subsystems $S=\{A_1,\ldots,A_n \}$,
the $n$-qubit W-class state (\ref{nWclass2}) can be also considered as an $m$-partite W-class state with different names
of the basis, and thus we can have following lemma.

\begin{Lem}\label{Lem:Wclass}
 For any $n$-qubit W-class states $\ket{W}_{A_1\cdots A_n}$ and a partition $P=\{P_1,\ldots,P_m \}$
of the set of subsystems $S=\{A_1,\ldots,A_n \}$,
\begin{equation}
\mathcal{C}_{P_s(P_1\cdots\widehat{P}_s\cdots P_m)}^2 = \sum_{k \neq s}\mathcal{C}_{P_s P_k}^2 = \sum_{k \neq s}(\mathcal{C}_{P_s P_k}^a)^2,
\label{eq:nbitpart}
\end{equation}
and
\begin{equation}
\mathcal{C}_{P_s P_k}=(\mathcal{C}_{P_s P_k}^a),
\end{equation}
for all $k \neq s$.
\end{Lem}

\subsection{$n$-qudit W-Class States}\label{Subsec:ndW}
Now, we generalize the concept of W-class states to arbitrary $n$-qudit systems with similar properties
of monogamy relations as in Lemma \ref{Lem:Wclass}.

Let us consider a class of $n$-qudit quantum states,
\begin{equation}
\ket{W_n^d}_{A_1\cdots A_n} = \sum_{i=1}^{d-1}(a_{1i} \ket{i\cdots 0}_{A_1\cdots A_n}+\cdots+a_{ni} \ket{0\cdots i}_{A_1\cdots A_n}),
\label{ndWclass}
\end{equation}
with $\sum_{s=1}^{n}{\sum_{i=1}^{d-1}{|a_{si}|^2}}=1$.
In the case $d=2$, Equation (\ref{ndWclass}) is reduced to $n$-qubit W-class states in (\ref{nWclass2}).

Let $|\tilde{\psi}_s\rangle_{A_s}=\sum _{i=1}^{d-1}a_{si}\ket{i}_{A_s}$ be an unnormalized state of subsystem $A_s$
for $s\in\{1,\ldots,n \}$
and $|\langle\tilde{\psi}_s|\tilde{\psi}_s\rangle|=\sum_{i=1}^{d-1}|a_{si}|^2 = \alpha_s^2$, then
 the normalization condition can be rephrased as $\sum_{s=1}^{n}\alpha_s^2 =1$.

Now, we will see that the class of states~(\ref{ndWclass}) has similar properties as
in Equation (\ref{nsatu}); that is,
\begin{eqnarray}
\mathcal{C}_{A_1 (A_2 \cdots A_n)}^2 &=& \mathcal{C}_{A_1 A_2}^2+\cdots+\mathcal{C}_{A_1 A_n}^2 ,
\nonumber\\
\mathcal{C}_{A_1 A_t}&=& \mathcal{C}_{A_1 A_t}^a,~~t = 2,\ldots, n.
\label{ndsatu}
\end{eqnarray}
In other words, CKW and its dual inequalities are saturated by the class of states~(\ref{ndWclass}). To see this, let us first consider $\mathcal{C}^2_{A_1 (A_2 \cdots A_n)}=2(1-\T\rho_{A_1}^2)$
where $\rho_{A_1}=\T_{A_2\cdots A_n}({\ket{W_n^d}_{A_1\cdots A_n}\bra{W_n^d}})$.
Since $\rho_{A_1}=|\tilde{\psi}_{1}\rangle_{A_1}\langle\tilde{\psi}_{1}|+\sum_{s=2}^{n} \alpha_{s}^2 \ket{0}_{A_1}\bra{0}$,
\begin{equation}
\mathcal{C}^2_{A_1 (A_2 \cdots A_n)}= 4\alpha_1^2(\sum_{s=2}^{n}\alpha_s^2).
\label{dC_1(2_n)}
\end{equation}

For $\mathcal{C}^2_{A_1 A_t}$ with $t \in \{2,\ldots,n\}$, let us consider
\begin{eqnarray}
\rho_{A_1 A_t} &=&\T_{A_2\cdots\widehat{A}_t\cdots A_n}({\ket{W_n^d}_{A_1\cdots A_n}\bra{W_n^d}})\nonumber\\
               &=&\sum_{i,j=1}^{d-1}(a_{1i}\ket{i0}+a_{ti}\ket{0i})_{A_1 A_t}(a^*_{1j}\bra{j0}+a^*_{tj}\bra{0j})\nonumber\\
               &&+(\alpha_2^2 +\cdots+\widehat{\alpha}_t^2+\cdots+\alpha_n^2)\ket{00}_{A_1 A_t}\bra{00},
\label{drhoA1At}
\end{eqnarray}
where $A_2\cdots\widehat{A}_i\cdots A_n = A_2\cdots A_{i-1}A_{i+1}\cdots A_n$ and
 $(\alpha_2^2 +\cdots+\widehat{\alpha}_t^2+\cdots+\alpha_n^2)=(\alpha_2^2 +\cdots+\alpha_{t-1}^2+\alpha_{t+1}^2+\cdots+\alpha_n^2)$.

Now, let us denote $\ket{\tilde{x}}_{A_1 A_t}=\sum_{i=1}^{d-1}(a_{1i}\ket{i0}+a_{ti}\ket{0i})_{A_1 A_t}$ and $\ket{\tilde{y}}_{A_1 A_t}=\sqrt{\alpha_2^2 +\cdots+\widehat{\alpha}_t^2+\cdots+\alpha_n^2}\ket{00}_{A_1 A_t}$,
where $\ket{\tilde{x}}_{A_1 A_t}$ and $\ket{\tilde{y}}_{A_1 A_t}$ are unnormalized states of the subsystems $A_1 A_t$.
Then, by the HJW theorem, any other decomposition of $\rho_{A_1 A_t}=\sum_{h=1}^{r}|\tilde{\phi}_h\rangle_{A_1 A_t}\langle\tilde{\phi}_h|$ with size $r \geq 2$
can be obtained by an $r\times r$ unitary matrix $(u_{hl})$ where
\begin{equation}
|\tilde{\phi}_h\rangle_{A_1 A_t}=u_{h1}\ket{\tilde{x}}_{A_1 A_t}+u_{h2}\ket{\tilde{y}}_{A_1 A_t}.
\label{ddecomp}
\end{equation}
Let $\rho_{A_1 A_t}=\sum_{h}{ p_h \ket{\phi_h}_{A_1 A_t}\bra{\phi_h}}$
where $\sqrt{p_h}\ket{\phi_h}_{A_1 A_t}=|\tilde{\phi}_h\rangle_{A_1 A_t}$ and $ p_h =|\langle\tilde{\phi}_h|\tilde{\phi}_h\rangle|$;
then, after a tedious calculation, we can see that the average concurrence of the pure state
decomposition $\rho_{A_1 A_t}=\sum_{h}{ p_h \ket{\phi_h}_{A_1 A_t}\bra{\phi_h}}$ is
\begin{eqnarray}
\sum_{h=1}^{r} p_h \mathcal{C}(\ket{\phi_h}_{A_1 A_t}) &=& \sum_{h=1}^{r}p_h \sqrt{2(1-\T{(\rho^h_{A_1})}^2)}\nonumber\\
 &=& 2\alpha_1 \alpha_t,
\label{dC1t}
\end{eqnarray}
where $\rho^h_{A_1} = \T_{A_t}(\ket{\phi_h}_{A_1 A_t}\bra{\phi_h})$.
Similar to the $n$-qubit case, the average concurrence remains the same for any pure state decomposition of $\rho_{A_1 A_t}$, and
thus, we can have,
\begin{eqnarray}
\min \sum_{h} p_h \mathcal{C}(\ket{\phi_h}_{A_1 A_t}) = \max \sum_{h} p_h \mathcal{C}(\ket{\phi_h}_{A_1 A_t}),
\label{davea1atsame}
\end{eqnarray}
where the maximum and minimum are taken over all possible decomposition of $\rho_{A_1 A_t}$.
This implies
\begin{equation}
\mathcal{C}_{A_1 A_t} = \mathcal{C}_{A_1 A_t}^a = 2\alpha_1 \alpha_t,
\label{dCa1atsame}
\end{equation}
which leads to (\ref{ndsatu}).

Now, let us consider a partition $P=\{P_1,\ldots,P_m \}$, $m\leq n$ where each of $P_s$ with $s \in \{1,\ldots,m\} $ contains several qudits
such that $|P_s|=n_s$ and $n_1 + \cdots + n_m =n$.
Without loss of generality, we may assume $P_1 =\{A_1, \ldots ,A_{n_1}\}$,
$P_2 = \{A_{n_1+1 },\ldots,A_{n_1+n_2} \}$,$\ldots$, $P_m = \{A_{n_1+\cdots+n_{m-1} +1 },\ldots,A_{n} \}$;
otherwise we can have some proper reordering of the subsystems.
For each party $P_s$ of the partition $P$, let
\begin{eqnarray}
|\tilde{x}_{si}\rangle_{P_s}&=&a_{(n_1+\cdots+n_{s-1}+1)i}\ket{i0\cdots 0}_{P_s}+ a_{(n_1+\cdots+n_{s-1}+2)i}\ket{0i\cdots 0}_{P_s}\nonumber\\
 &&+\cdots+a_{(n_1+\cdots+n_s) i}\ket{00\cdots i}_{P_s},
\label{PsWi}
\end{eqnarray}
then $|\tilde{x}_{si}\rangle_{P_s}$ is an unnormalized state of the party $P_s$ and (\ref{ndWclass}) can be rewritten as
\begin{eqnarray}
\ket{W_n^d}_{P_1\cdots P_m}=\sum_{i=1}^{d-1}&& \left( |\tilde{x}_{1i}\rangle_{P_1}\otimes|\vec{0}\rangle_{P_2}\otimes\cdots\otimes|\vec{0}\rangle_{P_m}\right.\nonumber\\
           && \left.+\cdots +|\vec{0}\rangle_{P_1}\otimes|\vec{0}\rangle_{P_2}\otimes\cdots\otimes|\tilde{x}_{mi}\rangle_{P_m}\right),
\label{dmpartW}
\end{eqnarray}
where $|\vec{0}\rangle_{P_s} = \ket{0\cdots0}_{P_s}$.
If we consider the normalized state $\ket{x_{si}}_{P_s} = \frac{1}{\sqrt{q_{si}}}|\tilde{x}_{si}\rangle_{P_s}$ with
$|\langle\tilde{x}_{si}|\tilde{x}_{si}\rangle|=q_{si}^2$ and rename $\ket{x_{si}}_{P_s}=\ket{i}_{P_s}$ and $|\vec{0}\rangle_{P_s}= \ket{0}_{P_s}$,
then (\ref{dmpartW}) can be represented as
\begin{eqnarray}
\ket{W_n^d}_{P_1\cdots P_m}=\sum_{i=1}^{d-1}&&\left(\sqrt{q_{1i}}\ket{i}_{P_1}\otimes\ket{0}_{P_2}\otimes\cdots\otimes\ket{0}_{P_m}\right. \nonumber\\
            &&\left.+\cdots+\sqrt{q_{mi}} \ket{0}_{P_1}\otimes\ket{0}_{P_2}\otimes\cdots\otimes\ket{i}_{P_m}\right),
\label{dmpartW2}
\end{eqnarray}
which is an $m$-partite generalized W-class state and, thus, we can have the second lemma which incorporates Lemma~\ref{Lem:Wclass}.

\begin{Lem}\label{Lem:dWclass}
 For any $n$-qudit generalized W-class states $\ket{W}_{A_1 \cdots A_n}$ in~(\ref{ndWclass}) and a partition $P=\{P_1, \ldots,P_m \}$
for the set of subsystems $S=\{A_1,\ldots,A_n \}$,
\begin{equation}
\mathcal{C}_{P_s(P_1 \cdots\hat{P}_s\cdots P_m)}^2 = \sum_{k \neq s}\mathcal{C}_{P_s P_k}^2 = \sum_{k \neq s}(\mathcal{C}_{P_s P_k}^a)^2,
\label{eq:ndbitpart}
\end{equation}
and
\begin{equation}
\mathcal{C}_{P_s P_k}=(\mathcal{C}_{P_s P_k}^a),
\end{equation}
for all $k \neq s$.
\end{Lem}
Furthermore, if we consider the state $|\tilde{x}_s\rangle_{P_s}$ of the partition $P_s$ such that
\begin{eqnarray}
|\tilde{x}_s\rangle_{P_s} = \sum_{i=1}^{d-1}&&( a_{(n_1+\cdots+n_{s-1}+1)i}\ket{i0\cdots 0}_{P_s}
+a_{(n_1+\cdots+n_{s-1}+2)i}\ket{0i\cdots 0}_{P_s}\nonumber\\
 &&+\cdots+a_{(n_1+\cdots +n_s)i}\ket{00\cdots i}_{P_s}),
\label{PsW}
\end{eqnarray}
then $|\tilde{x}_s\rangle_{P_s}$ is an unnormalized W-class state of $n_s$-qudit system.
Let $|\tilde{x}_s\rangle_{P_s}=\sqrt{q_s}\ket{W_{n_s}^d}_{P_s}$ with $q_s^2 = \sum_{i=1}^{d-1}q_{si}^2$; then (\ref{dmpartW}) can be
also represented as
\begin{eqnarray}
\ket{W}_{P_1\cdots P_m} &=& \sqrt{q_1}\ket{W_{n_1}^d}_{P_1}\otimes \ket{0}_{P_2}\otimes\cdots\otimes \ket{0}_{P_m}\nonumber\\
                         &&+ \cdots+ \sqrt{q_m}\ket{0}_{P_1}\otimes \ket{0}_{P_2}\otimes\cdots\otimes\ket{W_{n_m}^d}_{P_m},
\label{dWinW}
\end{eqnarray}
which is the same type of representation as the $n$-qubit W-class states in (\ref{2WinW}).

\section{$n$-qudit Mixed States}\label{Sec:mixed}
Concurrence is one of most well-known entanglement measures for bipartite quantum system
with an explicit formula for $2$-qubit system.
However, in higher-dimensional quantum system,  there does not exist any explicit way of
evaluation yet for mixed state. The lack of an analytic evaluation technique is mostly due to the difficulty of optimization problem which is minimizing over all possible pure state decompositions of the given mixed state. The difficulty of optimization problem for its dual, CoA, also arises in forms of maximization.

Recently, the optimal pure state decomposition for the mixture of generalized GHZ and W states in $3$-qubit system was found~\cite{LOSU,EOSU},
and the optimal decomposition, here, was assured by the saturation of CKW inequality.
In~\cite{FOR}, another monogamy relation in terms of the higher-tangle, the squared concurrence of  pure states and
its convex-roof extension for mixed states, was also investigated for a mixture of $n$-qubit W-class state and a product state
$\ket{0}^{\otimes n}$.

Here, we investigate the structure of the mixed states that consists of $n$-qubit W-class states and the product state
$\ket{0}^{\otimes n}$ and provide an analytic proof for its saturation of CKW and its dual in equality.
This saturation of the inequalities are also true for any partition of the set of subsystems.
Noting that the average of squared concurrences is always an upper bound of the square of average concurrences,
the result in~\cite{FOR} becomes a special case of the result here.

For any $n$-qudit W-class state in (\ref{ndWclass}), if we consider the reduced density matrix of the subsystem
$\{A_{s_1},\ldots,A_{s_l}\}$ for $2 \leq l \leq  n-1 $, we can easily check that it is always a mixture of
some $l$-qudit W-class state and a product state $\ket{0\cdots0}$, that is,
\begin{equation}
\rho_{ A_{s_1}\cdots A_{s_l}} = p \ket{W_l^d}_{A_{s_1}\cdots A_{s_l}}\bra{W_l^d} + (1-p)\ket{0\cdots0}_{A_{s_1}\cdots A_{s_l}}\bra{0\cdots0},
\label{mixed1}
\end{equation}
for some $0 \leq p \leq 1$.

Conversely, let us consider any mixture of a $n$-qudit W-class state in (\ref{ndWclass}) and a product state $\ket{0\cdots0}_{A_1\cdots A_n}$,
\begin{equation}
\rho_{A_1\cdots A_n} = p \ket{W_n^d}_{A_1\cdots A_n}\bra{W_n^d} + (1-p)\ket{0\cdots0}_{A_1...A_n}\bra{0\cdots0}.
\label{mixed2}
\end{equation}
Since $\rho_{A_1\cdots A_n}$ is an operator of rank two, we can always have a purification $\ket{\psi}_{A_1\cdots A_n A_{n+1}} \in~(\mathbb{C}^d)^{\otimes n+1}$ of $\rho_{A_1\cdots A_n}$ such that,
\begin{eqnarray}
\ket{\psi}_{A_1\cdots A_n A_{n+1}}&=&\sqrt{p}\ket{W_n^d}_{A_1\cdots A_n}\otimes\ket{0}_{A_{n+1}}\nonumber\\
&&+\sqrt{1-p}\ket{0\cdots 0}_{A_1\cdots
A_n}\otimes\ket{x}_{A_{n+1}}, \label{puri}
\end{eqnarray}
where $\ket{x}_{A_{n+1}}=\sum_{1=i}^{d-1}a_{n+1i}\ket{i}_{A_{n+1}}$ is a $1$-qudit quantum state of $A_{n+1}$
which is orthogonal to $\ket{0}_{A_{n+1}}$ where $\sum_{1=i}^{d-1}|a_{n+1i}|^2 =1$.
Now, we can easily see that (\ref{puri}) can be rewritten as
\begin{eqnarray}
\ket{\psi}_{A_1\cdots A_{n+1}}= \sum_{i=1}^{d-1}&&[\sqrt{p}(a_{1i}\ket{i\cdots00}_{A_1\cdots A_{n+1}}+\cdots+a_{ni} \ket{0\cdots i0})_{A_1\cdots A_{n+1}}\nonumber\\
&&+ \sqrt{1-p}a_{n+1i}\ket{0\cdots 0i}_{A_1\cdots A_{n+1}}], \label{puri2}
\end{eqnarray}
and this is an $(n+1)$-qudit W-class state.

In other words, the reduced density matrix of a generalized W-class state onto any subsystem is a mixture of a W-class state
and a product state. Furthermore, any mixture of a W-class state and a product state $\ket{0\cdots0}$ can be considered
 as a reduced density matrix of some W-class state in a quantum system with a larger number of parties.

Then, we can have following theorem.
\begin{Thm}\label{Thm:dWclass}
Let $\rho_{A_1\cdots A_n}$ be a $n$-qudit mixed state in $\mathcal{B}((\C^d)^{\otimes n})$,
which is a mixture of a generalized W-class state $\ket{W}_{A_1 \cdots A_n}$ and
a product state $\ket{0\cdots 0}_{A_1 \cdots A_n}$ with any weighting factor $0\leq p \leq 1$ such that
\begin{equation}
\rho_{A_1\cdots A_n} = p \ket{W_n^d}_{A_1\cdots A_n}\bra{W_n^d} + (1-p)\ket{0\cdots 0}_{A_1\cdots A_n}\bra{0\cdots 0}.
\label{mixed3}
\end{equation}
Then
\begin{eqnarray}
\mathcal{C}_{A_1 (A_2 \cdots A_n)}^2 &=& \mathcal{C}_{A_1 A_2}^2+\cdots+\mathcal{C}_{A_1 A_n}^2 ,
\nonumber\\
\mathcal{C}_{A_1 A_t}&=& \mathcal{C}_{A_1 A_t}^a,~~t = 2,\ldots,n.
\label{mixsatu}
\end{eqnarray}
Furthermore, for any partition $P=\{P_1,\ldots,P_m \}$ of the set of subsystems $S=\{A_1,\ldots,A_n \}$,
\begin{equation}
\mathcal{C}_{P_s(P_1 \cdots\hat{P}_s\cdots P_m)}^2 = \sum_{k \neq s}\mathcal{C}_{P_s P_k}^2 = \sum_{k \neq s}(\mathcal{C}_{P_s P_k}^a)^2,
\label{eq:ndbitpart}
\end{equation}
and
\begin{equation}
\mathcal{C}_{P_s P_k}=(\mathcal{C}_{P_s P_k}^a),
\end{equation}
for all $k \neq s$.
\end{Thm}

\begin{proof}
Let (\ref{puri2}) be a purification of $\rho_{A_1\cdots A_n}$ in $(\mathbb{C}^d)^{\otimes n+1}$,
then, by Lemma \ref{Lem:dWclass}, we can have,
\begin{eqnarray}
\mathcal{C}^2_{A_1[(A_2\cdots A_n)A_{n+1}]}&=&\mathcal{C}^2_{A_1(A_2\cdots A_n)} + \mathcal{C}^2_{A_1 A_{n+1}}\nonumber\\
&=&\sum_{s=2}^{n}\mathcal{C}^2_{A_1A_s} + \mathcal{C}^2_{A_1 A_{n+1}},
\label{n+1satu}
\end{eqnarray}
and, thus,
\begin{equation}
\mathcal{C}^2(\rho_{A_1(A_2\cdots A_n)})=\mathcal{C}^2_{A_1(A_2\cdots A_n)} = \sum_{s=2}^{n}\mathcal{C}^2_{A_1A_s}.
\label{mixsatu}
\end{equation}

Furthermore, for any partition $P=\{P_1,\ldots,P_m \}$ of the set of subsystems $S=\{A_1,\ldots,A_n \}$,
\begin{eqnarray}
\mathcal{C}^2_{P_i[(P_2 \cdots\hat{P}_i\cdots P_m) A_{n+1}]}&=&\mathcal{C}^2_{P_i(P_2 \cdots\hat{P}_i\cdots P_m)}
 + \mathcal{C}^2_{P_i A_{n+1}}\nonumber\\
&=&\sum_{s \neq i}\mathcal{C}^2_{P_iP_s} + \mathcal{C}^2_{P_i A_{n+1}},
\label{n+1satu2}
\end{eqnarray}
and we can have
\begin{equation}
\mathcal{C}^2_{P_i (P_2\cdots\hat{P}_i\cdots P_m)}= \sum_{s \neq i}\mathcal{C}^2_{P_iP_s}.
\label{mixPsatu}
\end{equation}
\end{proof}

Note that Theorem~\ref{Thm:dWclass} encapsulates the first two lemmas. In other words, if $p=1$, Theorem~\ref{Thm:dWclass} deals with the
monogamy relations of $n$-qudit W-class states that were presented in Lemma~\ref{Lem:dWclass}, and for the case $p=1,~d=2$, it is about
the monogamy relations of $n$-qubit W-class states that were presented in Lemma~\ref{Lem:Wclass}. Thus, Theorem~\ref{Thm:dWclass} deals with
the most general case of high-dimensional multipartite mixed states, so far, whose entanglement is completely characterized by their
partial entanglements.

\section{Conclusions}\label{Sec:conclusion}
We have investigated general monogamy relations for W-class states by considering the structure of W-class states in terms of arbitrary partitions of subsystems. We have generalized the concept of W-class states from $n$-qubit systems to arbitrary $n$-qudit systems and
have shown that their entanglement is completely characterized by their partial entanglements in terms of any partition of the set of subsystems.
The structural properties of a class of mixed states that consist of a $n$-qubit W-class state
and a product state $\ket{0}^{\otimes n}$ have been shown, and we have proved that the monogamy relations of the mixture have the same
characterization as the case of W-class states.

The structural properties of W-class states by considering an arbitrary partition of the set of subsystems show that
the structure of W-class states is inherent with respect to any arbitrary partition by the choice of a proper basis.
Our novel technique can also be helpful to study the structural properties of other genuine multipartite entangled states, such as
$n$-qudit GHZ-class states and cluster states~\cite{BR}. Noting the importance of the study on high-dimensional multipartite entanglement, although there have not been much so far, our results can provide a rich reference for future work on the characterization of multipartite entanglement.

\section*{Acknowledgments}
This work was supported by the Korea Research Foundation Grant funded by the Korean Government (MOEHRD) (KRF-2007-357-C00008),
Alberta's informatics Circle of Research Excellence (iCORE) and a CIFAR Associateship.

\end{document}